\documentclass[twocolumn]{aastex631}
\usepackage{amsmath}

\newcommand\kms{km~s$^{-1}$}
\newcommand\msun{M$_\odot$}

\newcommand{\rev}[2]{#2}
 \newcommand{\revv}[2]{#2}

\begin{document}

\title{Threading the Magellanic Needle:\\Hypervelocity Stars Trace the Past Location of the LMC}

\author[0000-0001-9982-0241]{Scott Lucchini}
\altaffiliation{Authors contributed equally}
\affiliation{Center for Astrophysics $|$ Harvard \& Smithsonian, 60 Garden Street, Cambridge, MA 02138, USA}

\correspondingauthor{Scott Lucchini}
\email{scott.lucchini@cfa.harvard.edu}

\author[0000-0002-6800-5778]{Jiwon Jesse Han}
\altaffiliation{Authors contributed equally}
\affiliation{Center for Astrophysics $|$ Harvard \& Smithsonian, 60 Garden Street, Cambridge, MA 02138, USA}

\begin{abstract}

Recent discoveries have shown that a \rev{significant }{}population of hypervelocity stars (HVSs) originate from the Large Magellanic Cloud (LMC). We use three such HVSs as dynamical tracers to constrain the past orbit of the LMC. Since each star was ejected at a finite time in the past, it must intersect the past position of the \rev{LMC}{LMC's central black hole} at its ejection time. We model the LMC's orbit under the influence of dynamical friction and extended mass distributions for both the LMC and the Milky Way, generating a large ensemble of orbital realizations. By evaluating which orbits intersect the back-integrated HVS trajectories, we compute posterior distributions over the LMC's orbital history. This approach provides significantly tighter constraints on the past motion of the LMC than previously possible.
We find two previously published orbital models that are consistent with these new constraints: a first-passage trajectory from a self-consistent hydrodynamic simulation, and a second-passage trajectory from a collisionless $N$-body simulation.
In parallel, we infer the present-day ejection site of the HVSs—likely tracing the LMC's dynamical center and supermassive black hole—independent of conventional methods.

\end{abstract}

\keywords{Galaxy dynamics (591); Large Magellanic Cloud (903)}

\section{Introduction} \label{sec:intro}



The Magellanic Clouds have long stood as the Milky Way’s most conspicuous and massive companions, and the Large Magellanic Cloud (LMC) in particular has played a starring role in our understanding of satellite–host dynamics. Early models, based on dynamical arguments and analytic potentials \citep[e.g.][]{Tremaine1976,Weinberg2000}, favored a picture in which the LMC was a long-term, many-passage satellite of the Milky Way (MW). However, this view was dramatically reshaped by the advent of precise proper motion measurements from the Hubble Space Telescope \citep{kallivayalil06,kallivayalil13}, which revealed that the LMC's tangential velocity is likely high enough that it is on its first passage around our Galaxy \citep{besla07}.

The first-passage scenario was not only consistent with the LMC’s present-day velocity, but also helped explain several puzzles in the Magellanic system, including the lack of extensive tidal disruption \citep{choi22,jimenez-arranz24} and the survival of the LMC’s circumgalactic medium (the ``Magellanic Corona''; \citealt{Lucchini2020,DK22,mishra24}). More recently, however, \citet{vasiliev24} showed that the current position and velocity of the LMC can also be reconciled with a more extended, bound orbit in which the LMC completed a previous pericentric passage roughly 6$-$8~Gyr ago, at a distance of $\sim$100~kpc. This re-opens the door to older scenarios, but with new constraints and numerical sophistication.

All of these models—first passage or second, fast or slow—share a common framework: they use the LMC's present-day position, velocity, and mass, to determine the viable orbits. Whether using analytic orbit integration \citep[e.g.][]{Patel2017,zivick18} or live $N$-body modeling \citep[e.g.][]{vasiliev24,sheng24}, increasingly precise observations and increasingly sophisticated simulations have brought the field to a new level of rigor. But fundamentally, they remain extrapolations into the past from the present.

Here, we present a new method that constrains the past location of the LMC directly. So-called hypervelocity stars (HVSs) are thought to originate from dynamical ejections involving a supermassive black hole (BH) and a binary system \citep{Hills1988}. Recently, \citet{Han2025} identified a \rev{substantial }{}population of HVSs whose trajectories trace back not to the Galactic center, but to the LMC itself. The ejection times of these stars span $\sim$30$–$400 Myr ago. By integrating the orbits of these stars back in time, we can infer the position of the LMC at the time of their ejection—effectively identifying a ``needle’s eye'' that the true orbit of the LMC must thread.

This makes the HVS population the first dynamical tracer that constrains the past position of the LMC. The resulting constraint complements traditional methods and offers new insight into the dynamical history of the MW–LMC system, in particular in distinguishing the first, second, or many passage scenario.

Another open question is the present-day location of the LMC's supermassive black hole that produced the HVSs. At present day, the black hole seems to be relatively inactive, and given the LMC’s tumultuous history, it may be offset from the galaxy’s center—which itself is only weakly constrained from observations \cite[e.g.][]{vandermarel14, Wan2020,Niederhofer22}. While the ejection location and time are moderately degenerate for any single HVS, this degeneracy can be broken by combining constraints from multiple HVSs. In doing so, we also gain insight into the current position of the black hole. 

In this Letter, we analyze three LMC-origin HVSs to derive new constrains on the past orbit of the LMC, and the present location of its supermassive black hole.

\section{Methods}\label{sec:methods}


\citet{Han2025} analyzed the full HVS Survey dataset \citep{Brown2014}, which includes 21 unbound late B-type main sequence stars with well-measured atmospheric parameters such as effective temperature, surface gravity, and projected rotational velocity ($v\sin i$). From this sample, \citet{Han2025} identified several HVSs that likely originated from the LMC. In this study, we focus on HVS~3, HVS~7, and HVS~15—the three stars with the highest likelihood of LMC origin, based on orbits that are incompatible with a Galactic center ejection. HVS~3, in particular, has a long-standing association with the LMC \citep{Gualandris2007, Bonanos2008, Przybilla2008, Erkal2019}. \rev{}{Throughout this work, we assume that these HVSs have originated from the LMC. Thus, we rule out any LMC orbits that are incompatible with the HVS trajectories.} While additional HVSs may also be consistent with an LMC origin, their positions and velocities are less well constrained than those of the three stars we analyze here. For further details on the HVS dataset, we refer the reader to \citet{Han2025}.
\rev{In the following section, we describe the orbit integration methods applied to this sample.}{}

\rev{}{Our methodology begins by sampling the present-day properties of the LMC—RA, dec, distance, proper motion, and radial velocity—and transforming them into Galactocentric coordinates. We also sample from uncertainties in the heliocentric-to-Galactocentric coordinate transformation. For each hypervelocity star, we sample its present-day properties and integrate its orbit analytically, treating it as a test particle in the evolving LMC–Milky Way potential. We then evaluate the likelihood of each sampled LMC orbit based on how many HVS trajectories pass near the LMC's center. Finally, we combine this likelihood with priors on the LMC’s initial position (RA, Dec, distance) to compute a posterior probability for each orbit.}


\begin{deluxetable*}{lccccccc}
\tablecaption{Initial conditions of the LMC and HVSs used to integrate orbits.}
\label{tab:ICdata}
\tablehead{\colhead{Object} & \colhead{RA} & \colhead{Dec} & \colhead{Distance} & \colhead{$\mu_\alpha$} & \colhead{$\mu_\delta$} & \colhead{RV} & \colhead{Gaia source\_id} \\
         & (deg) & (deg) & (kpc) & (mas~yr$^{-1}$) & (mas~yr$^{-1}$) & (\kms) & }

 \startdata
LMC Center & $79.7\pm5.0$ & $-69.1\pm2.0$ & $50.1\pm2.5$ & $1.878\pm0.007$ & $0.293\pm0.018$ & $262.2\pm3.4$ & $-$ \\[1ex]
HVS 3 & 69.55 & -54.55 & $61\pm10$ & $0.851\pm0.11$ & $1.936\pm0.162$ & $723\pm3$ & 4777328613382967040 \\
HVS 7 & 173.30 & 1.14 & $52.17\pm6.25$ & $-0.09\pm0.18$ & $0.02\pm0.13$ & $526.9\pm3.0$ & 3799146650623432704 \\
HVS 15 & 173.42 & -1.35 & $66.16\pm9.75$ & $-1.30\pm0.36$ & $-0.48\pm0.23$ & $461.0\pm6.3$ & 3794074603484360704 \\
\enddata
\end{deluxetable*}

\subsection{Orbit Integration}

We use \texttt{gala}\footnote{\url{https://github.com/adrn/gala}} \citep{galajoss} with dynamical friction including the LMC and MW as live points with analytic, radially extended potentials while the HVSs are included as test particles. We generate 10,000 realizations of the LMC orbit, each with 10,000 Monte Carlo sampled orbits for each of the three HVSs (given their respective distance, proper motion, and radial velocity uncertainties). The initial conditions for the LMC Center and the three HVSs are listed in Table~\ref{tab:ICdata}. The spread in the LMC Center is intentionally large in Table~\ref{tab:ICdata}: the true dynamical center of the LMC is yet unknown \citep[e.g.][]{vandermarel02, vandermarel14, boyce17, Pietrzynski19, Wan2020, Niederhofer22}. Thus, we choose to sample the full space of possibilities within the LMC. Meanwhile, the systemic proper motion and radial velocities are well constrained, and we adopt values from \citet{Wan2020} and \citet{vandermarel02}.

We also take into account the uncertainty on the Solar position and velocity, which is required to transform the observed properties of the LMC and HVSs into Galactocentric coordinates. We take the Solar phase space coordinates and associated uncertainties from \citet{reid19}, but with the uncertainties increased by a factor of two to fully explore the phase space: $(U_\odot,V_\odot+\Theta_0,W_\odot)=(10.6\pm2.4,247.0\pm8.0,7.6\pm1.4)$~\kms, $R_0=8.15\pm0.3$~kpc, and $z_0=5.5\pm11.6$~pc. We then Monte Carlo sample the Solar phase space coordinates 10,000 times, each yielding a distinct transformation from heliocentric observations to Galactocentric coordinates. We implement these transformations in \texttt{astropy}  \citep{astropy:2013,astropy:2018,astropy:2022}.

Given a particular realization of the heliocentric-Galactocentric transformation, paired with an LMC Center initial condition, we simultaneously track the orbit of the LMC, the Milky Way (both galaxies are treated as live points with an extended mass distribution), and HVSs (10,000 Monte Carlo test particles per HVS).

\subsection{Potentials}

We use analytic, radially extended potentials for the LMC and MW, whose centers are allowed to accelerate in response to gravitational forces. For each galaxy, we use a superposition of a Miyamoto-Nagai disk potential \citep{miyamotonagai} and a dark halo potential (modeled by a Navarro-Frenk-White profile for the MW, \citealt{nfw96}; and a Hernquist profile for the LMC, \citealt{hernquist90}). For the MW, we use values based on \citet{bland-hawthorn16}: the disk has mass $3.6\times10^{10}$~\msun, scale length $a=2.5$~kpc, and scale height $b=500$~pc, and the halo has $M_{200}=1.1\times10^{12}$~\msun\ and concentration $c=10$. 
\revv{}{We note that doubling the Galactic disk mass alters the orbit of HVS~3 by $\sim0.1\%$, HVS~7 by $\sim20\%$, and HVS~5 by $\sim7\%$ compared to their spread induced by observational uncertainties. Given that these deflections are still subdominant to observational uncertainties$-$even in the unrealistic scenario of a disk twice as heavy as the literature value$-$we do not consider further variations to the disk parameters.}
For the LMC, we use $m=3.5\times10^9$~\msun, $a=1.5$~kpc, and $b=500$~pc for the disk, and $M=1.5\times10^{11}$~\msun and $a_s=15$~kpc for the Hernquist halo.

In addition to the fiducial simulations described above, we ran a set of test simulations where we substantially varied the MW and LMC dark matter halo masses. We found that the MW and LMC masses play a subdominant role in the orbital trajectories of HVSs. This can be credited to the fact that the HVSs are traveling so fast (several times the local escape velocity), and only integrate backwards up to $\sim$400~Myr before they reach the LMC. However, over longer timescales, the relative masses of the galaxies can effect the period of the LMC's orbit. The period is important, as it determines whether or not the LMC is on its first passage (i.e., period is smaller than Hubble time). However, other factors such as hydrodynamics and accretion history can play an equally significant role over these longer timescales. We discuss the effect of varying the LMC and MW masses in Appendix~\ref{app:masses}. For the main conclusions of the paper, we fix the mass of the LMC and the MW to the fiducial values.

\subsection{Dynamical Friction}

We model dynamical friction \citep{Chandrasekhar1943} following \citet{Patel2020} using
\begin{multline*}
\vec{F}_{df,\mathrm{MW}} = -\frac{4\pi G^2 M_\mathrm{sat}\ln \Lambda \rho(r)}{v^2}\times\\
\left[\mathrm{erf}(X)-\frac{2X}{\sqrt{\pi}}\exp(-X^2)\right]\frac{\vec{v}}{v}
\end{multline*}
where $M_\mathrm{sat}$ is the total mass of the LMC, $\rho(r)$ is the density of the MW's DM halo at radial distance $r$, $v$ is the velocity of the LMC relative to the MW, and $X=v/\sqrt{2\sigma}$ where $\sigma$ is the velocity dispersion which is calculated as a function of the maximum velocity in the galaxy's rotation curve following \citet{Zentner2003}. $\ln\Lambda$ is the Coulomb logarithm which, for the LMC moving through the MW, is determined following \citet{vanderMarel2012}:
\begin{equation*}
    \ln\Lambda_\mathrm{LMC}=\max\left(L,\ln(r/Ca_s)^\alpha\right)
\end{equation*}
where $L=0$, $C=1.22$, and $\alpha=1.0$ are the best fit values from \citet{Patel2017} and $a_s$ is the Hernquist scale length of the satellite.




\subsection{The Likelihood Function} \label{sec:likelihood}

We define the likelihood of a given LMC orbit (specified along with a Solar position and velocity) as the fraction of 10,000 Monte Carlo realizations of the HVS orbit that pass within a 1~kpc cube centered on the LMC Center. \rev{}{1~kpc was chosen so as to encapsulate the maximum region within which we would expect the BH to wander (see Section~\ref{sec:wandering}).} Each HVS thus yields a likelihood distribution over the space of possible LMC orbits. Of particular interest is the present-day position of the LMC Center. For each HVS, we construct a likelihood distribution for the Center's sky position, which we approximate as a two-dimensional Gaussian in right ascension and declination. To incorporate prior knowledge, we multiply this likelihood by a prior centered on the visual center of the LMC disk, (RA, Dec) $=(79.2,-68.7)$ degrees, with a full width at half maximum of $3.8^\circ=3.3$~kpc. 
This prior is intentionally broad—it serves only to loosely constrain the Center’s location, effectively excluding regions well outside the LMC disk. We also adopt a Gaussian prior on the LMC distance, centered at 50~kpc with a standard deviation of 3~kpc. The result is a posterior distribution over LMC orbits, including the present-day right ascension, declination, and distance of the LMC Center.
For the combined results, we first multiply the likelihoods for each star to obtain a combined likelihood. We then multiply the Gaussian prior to obtain the combined posterior distribution. We also tested with a tophat prior across the face of the LMC which we discuss in Appendix~\ref{app:prior}.

\begin{figure*}
    \centering
    \includegraphics[width=0.9\textwidth]{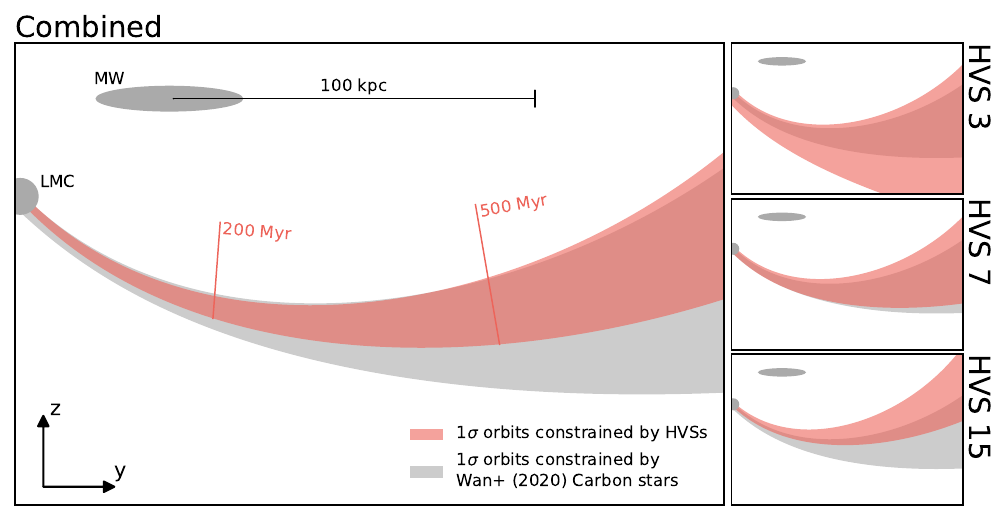}
    \caption{
    \rev{Orbital history of the LMC.}{Orbital history of the LMC, using \cite{Wan2020} measurements of the systemic proper motions.} The grey shaded region shows the $1\sigma$ extent of the extrapolated orbit based on the present-day dynamical center measured by \citet{Wan2020}, while the red shaded region shows the constraints derived from hypervelocity stars (HVSs). The HVS-based constraints are $\sim$2$\times$ tighter than those from simple extrapolation, despite allowing the LMC’s dynamical center to vary freely; the prior on the center spans a much larger region than the \citet{Wan2020} measurement. All HVSs shown were ejected less than 500 Myr ago. The main panel displays the combined constraints from multiple HVSs, while the side panels show individual constraints from HVS 3, HVS 7, and HVS 15.
    }
    \label{fig:orbits}
\end{figure*}

\section{Results}

The results of our analysis are twofold. First, we constrain the orbital history of the LMC using the trajectories of HVSs as dynamical tracers. As a natural corollary, we also obtain constraints on the present-day location of the LMC Center—the site where its central supermassive black hole is most likely to reside. We present both sets of results below.

\begin{figure*}
    \centering
    \includegraphics[width=0.85\textwidth]{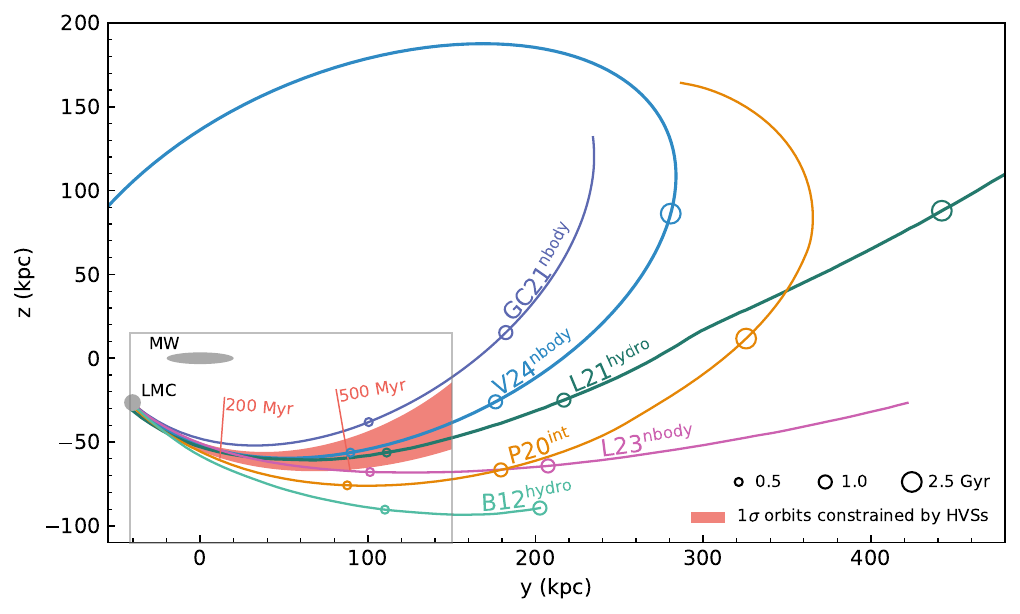}
    \caption{
    Comparison of the LMC’s past orbit as constrained in this study with previously published models. The red shaded region shows the $1\sigma$ range of trajectories allowed by HVSs. Each curve represents a different orbital model, with model types indicated in superscript: $^{\mathrm{nbody}}$ for $N$-body simulations \citep{garavito-camargo21, vasiliev24, lilleengen23}, $^{\mathrm{hydro}}$ for hydrodynamical simulations \citep{lucchini21, besla12}, and $^{\mathrm{int}}$ for analytic integrations \citep{Patel2020}. The grey inset box corresponds to the spatial region shown in Figure~\ref{fig:orbits}, or roughly $\sim$800~Myr into the past. Beyond this range, the timescales significantly predate the ejection of the HVSs and enter a regime where their trajectories no longer provide meaningful constraints. Among the models shown, the second passage $N$-body model of \citet{vasiliev24} and the first passage hydrodynamical model of \citet{lucchini21} are most consistent with the HVS-based constraints.
    }
    \label{fig:orbits-lit}
\end{figure*}

\subsection{The Orbital History of the LMC}

Figure~\ref{fig:orbits} displays the region of space traversed by the LMC over the past $\sim$800~Myr, as inferred from the trajectories of HVSs. The Milky Way is depicted as a grey oval, with the Sun located near $x \sim -8$~kpc. The red shaded area traces the $1\sigma$ region consistent with the combined constraints from multiple HVSs, under the assumption that each star was ejected from the center of the LMC.
\rev{}{Specifically, we plot the extent of orbital trajectories with posterior values above $e^{-1/2}$ times the maximum posterior value.}
The main panel shows the joint constraint from all available HVSs, while the side panels display the individual contributions from HVS~3, HVS~7, and HVS~15. For comparison, the grey shaded region in the main panel shows the result of integrating the LMC’s orbit backward in time from the present-day dynamical center measured by \citet{Wan2020}, using the same method described in Section~\ref{sec:methods}. Note that this method does not incorporate the systematic uncertainty associated with whether the adopted center truly represents the LMC’s dynamical center. Meanwhile, LMC's dynamical center is allowed to vary freely in the red shaded region. Despite this disadvantage, the HVS-based approach still yields a $\sim$2$\times$ tighter constraint on the LMC’s past trajectory. These constraints remain robust out to $\sim$800~Myr ago; beyond this, the extrapolations become increasingly sensitive to additional physics such as mass loss, dynamical friction, and hydrodynamics. We note that the LMC orbit is largely constrained to the $y-z$ plane as shown here; there is very little variation in the $x-$axis in either method.

In Figure~\ref{fig:orbits-lit}, we compare the LMC orbit constrained from HVSs to a selection of previously published orbital models. The red shaded region again shows the $1\sigma$ envelope derived from HVSs, while the overlaid curves represent individual orbits drawn from a variety of modeling approaches. These include $N$-body simulations ($^{\mathrm{nbody}}$; \citealt{garavito-camargo21}, \rev{}{MW 1 + LMC 3}; \citealt{vasiliev24}, \rev{}{$\mathcal{M}11-\mathcal{L}3$}; \citealt{lilleengen23}), hydrodynamical models ($^{\mathrm{hydro}}$; \citealt{lucchini21, besla12}), and analytic orbital integrations ($^{\mathrm{int}}$; \citealt{Patel2020}). Among the two hydrodynamical models, \citet{lucchini21} include the circumgalactic medium of the LMC (the ``Magellanic Corona''), which introduces additional friction to the LMC$-$MW system. Additionally, the orbit of \citet{besla12} used a fixed location for the MW which prevents the reflex motion of the MW towards the LMC. As a result, the LMC's approach is artificially shifted towards the $-z$ direction. The grey inset box denotes the same $\sim$800~Myr window highlighted in Figure~\ref{fig:orbits}, beyond which the HVSs no longer constrain the orbit due to their limited flight times. Among the various models, the second-passage $N$-body model of \citet{vasiliev24} and the first-passage hydrodynamical model of \citet{lucchini21} show the closest agreement to the HVS constraints. We further discuss these models in Section \ref{sec:discussion}.

\rev{}{We also constrain the ejection times of the HVSs from the LMC's central black hole. For HVSs 3, 7, and 15, we infer ejection times of $19 \pm 2$, $219^{+27}_{-40}$, and $299^{+36}_{-51}$ Myr ago, respectively. Our orbital integrations extend back 400 Myr, consistent with the broad age estimates for these main-sequence stars, and this upper limit comfortably encompasses our dynamical ejection time estimates. We find a correlation between the precision of the orbital constraint and the ejection time: the stars ejected further in the past (HVSs 7 and 15) yield more localized constraints on the LMC’s orbit, as seen in the smaller red regions in Figure~\ref{fig:orbits}. This suggests that identifying even older stars ejected from the LMC could further tighten the dynamical constraints on its past trajectory.
}

\subsection{The Present-Day Center of the LMC}

\begin{figure*}
    \centering
    \includegraphics[width=0.67\textwidth]{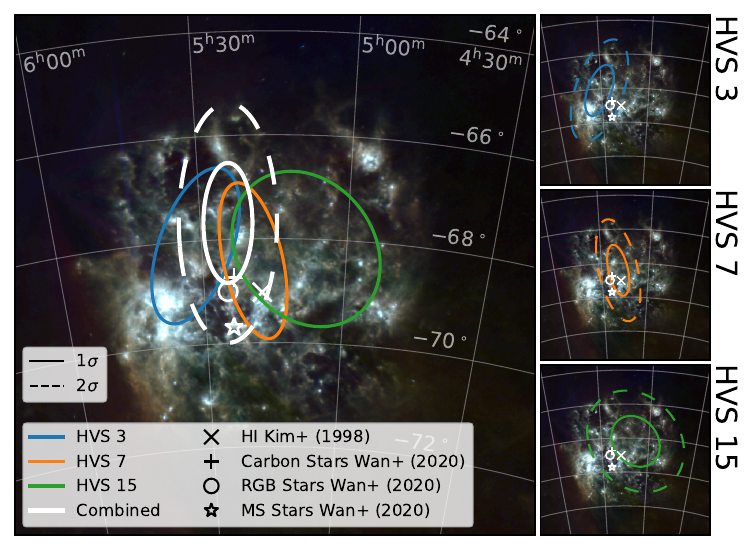}
    \caption{The on-sky distributions of the present-day positions of the LMC dynamical center that are consistent with the ejection of the hypervelocity stars. The right panels show the 1 (solid) and 2 (dashed) sigma curves for HVS 3, 7, and 15, compared against the dynamical centers determined using the \ion{H}{1} (cross), carbon stars (plus), red giant branch stars (circle), and main sequence stars (star) \citep{kim98,Wan2020}. The left panel shows the 1$\sigma$ contours for each star as well as the combined posterior in white.}
    \label{fig:face-on}
\end{figure*}

\rev{Figure~\ref{fig:face-on} shows the 1 and 2~$\sigma$ contours for the posterior distributions of the best-fit present-day position of the LMC's dynamical center. These positions correspond to the initial locations of the LMC in our analytic integrations, whose orbits are the most likely to have ejected each HVS.}{Each of our LMC orbits was generated by sampling from the distribution of RA and Dec listed in Table~\ref{tab:ICdata}. Thus we can map the orbit likelihoods and posteriors as a function of present-day RA and Dec of the center of the LMC's dark matter potential, which we expect to correlate well with the position of the LMC's SMBH (see discussion in Section~\ref{sec:wandering}). By fitting a 2D Gaussian to the distribution of posterior values as a function of RA and Dec, we can visualize where we expect the SMBH to be located. Figure~\ref{fig:face-on} shows the 1 and 2$\sigma$ contours for these Gaussian fits.}
Each of the hypervelocity stars provides a different constraint (right panels), and when combined (left panel) we see a tight posterior distribution for the location of the LMC's central BH. This distribution is best fit by a 2D Gaussian centered on (RA, Dec) = ($80.72^{\circ}$, $-67.79^{\circ}$), with $\sigma_x=0.44^\circ$, $\sigma_y=0.80^\circ$, and position angle $\theta=0^\circ$.

Figure~\ref{fig:face-on} also shows the locations of the LMC's dynamical center as determined by four different tracers in the literature. The white cross, plus, circle, and star denote the dynamical centers of the \ion{H}{1} \citep{kim98}, carbon stars, red giant branch stars, and main sequence stars \citep{Wan2020}. Each of these locations is offset by up to a degree from each other. Our predicted location for the LMC’s dynamical center—traced by its central black hole—is approximately 1.5$^\circ$ to the North compared to the average of previously measured centers. If the black hole resides at one of these centers, our results suggest that the carbon star–based center is the most consistent with the HVS constraints, while the center defined by main sequence stars is the least favored.

\section{Discussion} \label{sec:discussion}

\subsection{LMC Orbital Period} \label{sec:period}

We begin by examining the orbital history of the LMC. Although its trajectory is commonly categorized as first-passage, second-passage, or many-passage, the distinction between the first two scenarios is often subtle. For instance, trajectories that initially appear similar can diverge significantly depending on the physical processes included in the modeling, as illustrated in Figure~\ref{fig:orbits-lit}. Our orbital integrations incorporate dynamical friction and live, radially extended potentials. However, additional effects such as $N$-body components that undergo nonspherical potential deformations and mass loss, or hydrodynamic interactions that enhance frictional energy dissipation, can substantially modify the inferred orbital history. Tracing the LMC’s motion back $\sim$800~Myr, we find that our HVS-constrained orbits are consistent with two previously published models: the first-passage orbit with hydrodynamics from \citet{lucchini21}, and the second-passage orbit using live $N$-body halos from \citet{vasiliev24}.

In \citet{vasiliev24}, the orbit labeled $\mathcal{M}11-\mathcal{L}3$ features a previous pericentric passage around the Milky Way approximately 6.3 Gyr ago, at a distance of $\sim$100 kpc. At such early times, the Milky Way’s mass is expected to have been roughly half of its present-day value \citep{semenov24}. As one traces the orbit further back in time, this lower mass would result in a longer orbital period compared to models where the Milky Way’s mass is held fixed. Consequently, even this second-passage orbit is unlikely to reflect a meaningful past interaction with the Galaxy. In terms of present-day dynamical effects that are measurable---such as the reflex motion of the Milky Way’s disk or the gravitational wake in the stellar halo---this second-passage model is effectively indistinguishable from a first-passage scenario.

Interestingly, the orbital trajectory from \citet{lucchini21} also aligns well with our combined HVS posterior at late times. This orbit was derived from an $N$-body+hydrodynamic simulation that included both the Milky Way’s circumgalactic medium and the Magellanic Corona. The presence of these ambient gas components introduces additional friction, allowing the LMC to lose a substantial amount of orbital energy very rapidly as it reaches the inner MW. As a result, the LMC's trajectory deviates noticeably from those predicted by purely analytic or collisionless $N$-body models, allowing the LMC to approach the MW at higher initial velocity and along a more linear path.


\rev{}{Another aspect at play here is of course the choice of MW and LMC masses in these various models. Those trajectories that are more tightly wound have generally used slightly higher MW masses (as we also find in Appendix~\ref{app:masses}). For the models in Figure~\ref{fig:orbits-lit}, the MW masses used were 1.6, 1.5, 1.3, 1.1, 1.1, and $0.9\times10^{12}$~\msun\ for \citet{garavito-camargo21}, \citet{besla12}, \citet{vasiliev24}, \citet{Patel2020}, \citet{lucchini21}, and \citet{lilleengen23}, respectively. So while the MW mass plays a role, the combination of the agreement between the \citet{vasiliev24} and \citet{lucchini21} orbits within the past 500~Myr, and their divergence at earlier times indicates that the different physics included in $N$-body vs hydrodynamic simulations is also a significant factor.}

\subsection{Non-spherical Deformations of the Potential}

In this study, we have used spherical \rev{}{halo} potentials despite evidence that the DM halos of the MW and LMC have been shown to be non-spherical and even non-axisymmetric \citep[e.g.,][]{garavito-camargo19, han22,vasiliev23}. As described in Section \ref{sec:methods}, the deflections in the HVS trajectories due to these deformations are smaller than the \textit{Gaia} proper motion uncertainties \citep{boubert20}. Future work may explore this effect in further detail.


\begin{figure*}
    \centering
    \includegraphics[width=1.0\textwidth]{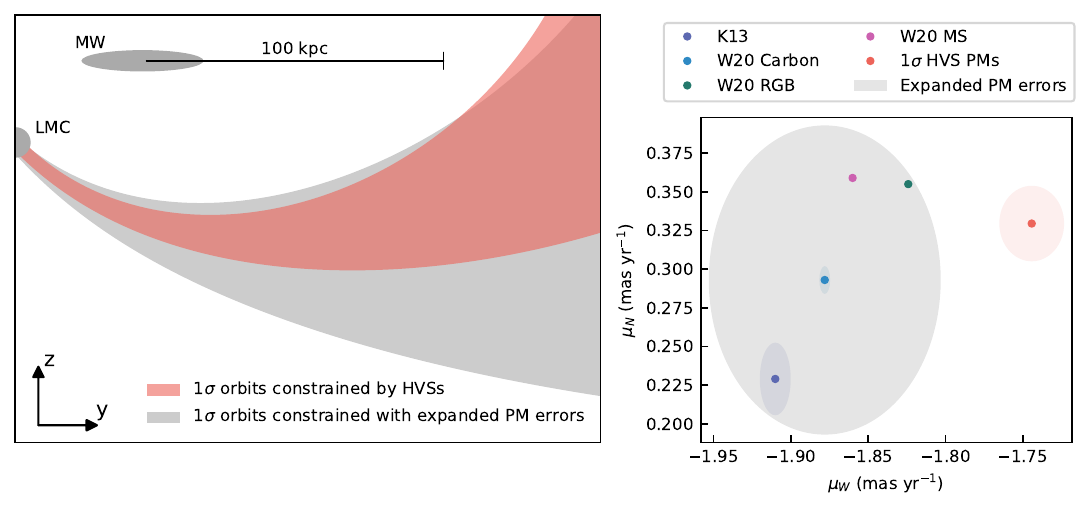}
    \caption{Orbital trajectory of the LMC with expanded proper motion errors. The left panel shows the combined available orbital space as in Figure~\ref{fig:orbits}, however the proper motion uncertainties have been artificially inflated so as to encompass the \citet{kallivayalil13} values. The grey region now shows the total 1$\sigma$ orbital space using the expanded RA, Dec, distance, and proper motion uncertainties, and the red region shows the region for those orbits with combined HVS posterior values above the 1$\sigma$ level. The right panel shows the proper motions from \citet{kallivayalil13} and \citet{Wan2020} in comparison to the expanded uncertainties used (grey oval) and the proper motions of the 1$\sigma$ HVS orbits (red dot and oval).}
    \label{fig:orbits_PMs}
\end{figure*}

\subsection{Is the BH at the Center of the LMC?} \label{sec:wandering}


In our orbit integrations, we track the center of the LMC's gravitational potential over time and compare it to the observed positions of the HVSs. This approach assumes that the LMC's central black hole resides at the potential minimum. However, this assumption may not strictly hold. Observational studies have shown that in low-mass galaxies ($M_* \leq 3 \times 10^9$~\msun), central black holes are often found to be ``wandering'' or offset from the dynamical center \citep[e.g.,][]{mezcua20,reines20}. Simulations support this behavior, attributing it to prior mergers and interactions with more massive halos \citep[e.g.][]{bellovary19,bellovary21}. Given that the LMC is currently undergoing a merger with the SMC and has recently entered the Milky Way’s halo, some displacement of its central black hole might be expected. However, because the SMC is significantly less massive than the LMC, and the LMC has only recently passed pericenter in its orbit around the Milky Way, we anticipate any such offset to be modest. Thus, the black hole is still likely to remain near the LMC’s potential minimum.

To test this assumption, we examine full $N$-body+hydrodynamic simulations of the evolution of the LMC, SMC, and Milky Way from \citet{lucchini21}. These simulations use GIZMO and include two interactions between the LMC and SMC over a 3.4 Gyr period, with a gravity softening length of 100 pc. The LMC begins 570 kpc from the Milky Way and evolves to its present-day position. After allowing the system to relax for 50 Myr, we compute the total gravitational potential of the LMC—including contributions from dark matter, stars, and gas—and identify the 50 most bound star particles. We then track their positions throughout the simulation as a proxy for the SMBH’s motion. The mean offset of these particles oscillates around zero with an amplitude of $\sim$100 pc, while the standard deviation increases gradually, reaching $\sim$200 pc after 3.5 Gyr. Importantly, the interactions with the SMC do not produce any discernible jumps in either the mean or the dispersion, suggesting that the location of the LMC’s most bound region—and by extension its SMBH—remains stable over time.

Thus, in this study, we assume that the LMC’s SMBH is located at the center of its gravitational potential, with a possible offset no larger than 500~pc, justified by the relatively recent ejection times of the HVSs ($<500$~Myr ago). Future work could examine this assumption more closely, potentially using the spatial and temporal distribution of HVS trajectories. If their ejection paths are found to cluster asymmetrically around the potential center at different epochs, this could provide valuable constraints on the motion of the SMBH relative to the LMC’s dynamical center.

\subsection{Effect of the SMC}

We also performed 3,000 orbit integrations that included the SMC, modeled with a mass of $10^{10}$~\msun. Over a 400 Myr timespan, 99.8\% of these integrations resulted in a single LMC–SMC interaction, with a median impact parameter of 9 kpc and a median interaction time of 114 Myr ago. Including the SMC introduces only modest changes to the inferred LMC orbit: the means of the likelihood distributions in initial right ascension, declination, and distance shift by $-0.1\%$, $0.4\%$, and $-4.7\%$, respectively, while the corresponding standard deviations change by $+9.2\%$, $-4.1\%$, and $-4.5\%$. These results suggest that while specific SMC trajectories might significantly enhance the likelihood of certain LMC orbits, their average effect on the overall orbital distribution is small. Future work can focus on identifying and studying these exceptional cases in greater detail.

\subsection{Systematic Uncertainty on the LMC proper motion}
\rev{}{While LMC proper motion measurements have a relatively small statistical uncertainty, there is a considerable spread among different tracers. To account for this systematic uncertainty, we make a fit where we inflate the proper motion uncertainties to include the \citet{Wan2020} values for all the tracers as well as the \citet{kallivayalil13} value. The right panel of Figure~\ref{fig:orbits_PMs} shows these proper motion values as well as our expanded uncertainties as the grey oval.}

\rev{}{Sampling from this wider distribution results in a larger orbital space available for the history of the LMC. In the left panel of Figure~\ref{fig:orbits_PMs}, the grey area denotes the LMC orbits possible with these expanded uncertainties (in RA, Dec, distance, and proper motion), while the red region shows the 1$\sigma$ region when taking the HVSs into consideration. As done in Figure~\ref{fig:orbits}, the red region is the full spread of all the orbits with posterior values (combined for all HVSs) above $e^{-1/2}$ times the maximum posterior value. We still find a $\sim$2$\times$ reduction in the orbital space.}

\section{Conclusion}

The discovery of hypervelocity stars originating from the LMC offers a unique opportunity to directly trace the LMC’s past orbital path. In this study, we employed a Bayesian framework to evaluate the likelihood of different LMC orbital trajectories based on their consistency with observed HVS paths. This approach yields the following key results:
\begin{enumerate}
\item The inclusion of HVS constraints reduces the viable parameter space of LMC orbital trajectories by a factor of two compared to trajectories obtained by backward integration from present-day properties, such as those reported by \citet{Wan2020} (Figure~\ref{fig:orbits}).
\item The inferred past trajectory of the LMC—based on HVSs 3, 7, and 15—is most consistent with the orbits proposed by \citet{vasiliev24} and \citet{lucchini21} (Figure~\ref{fig:orbits-lit}).
\item The LMC’s true dynamical center, presumed to host its central massive black hole, is localized to (RA, Dec) = ($80.72^{\circ}$, $-67.79^{\circ}$). This distribution is best described by a two-dimensional Gaussian with $\sigma_x = 0.44^\circ$, $\sigma_y = 0.80^\circ$, and position angle $\theta = 0^\circ$ (Figure~\ref{fig:face-on}).
\end{enumerate}

We emphasize that the constraints presented in this study are based on just three HVSs with currently available \textit{Gaia} DR3 proper motions. With improved proper motion measurements, these orbital constraints can be significantly tightened. From an observational standpoint, refining the proper motions of a small number of stars—e.g., with follow-up observations from the \textit{Hubble} or \textit{James Webb Space Telescope}—is a more tractable goal than surveying the entire LMC with extreme precision. We therefore propose that modeling the orbits of HVSs represents a promising and efficient path toward constraining the LMC’s orbital history and pinpointing the current location of its central supermassive black hole.

Finally, future work will explore the orbital space of the LMC with full hydrodynamic simulations in order to obtain additional constraints on the LMC's history as well as the SMC's properties and orbit by leveraging observational properties of the Magellanic Stream.\\

\noindent
The authors would like to acknowledge the many fruitful discussions at the XMC II: Clouds over Yellowstone conference that helped refine this work.
Support for SL was provided by Harvard University through the Institute for Theory and Computation Fellowship. The computations in this paper were run on the FASRC cluster supported by the FAS Division of Science Research Computing Group at Harvard University.

\software{gala \citep{galajoss,gala1.9.1}}

\bibliography{references}{}
\bibliographystyle{aasjournal}

\appendix

\section{MW and LMC Masses} \label{app:masses}

\begin{figure*}
    \centering
    \includegraphics[width=0.8775\textwidth]{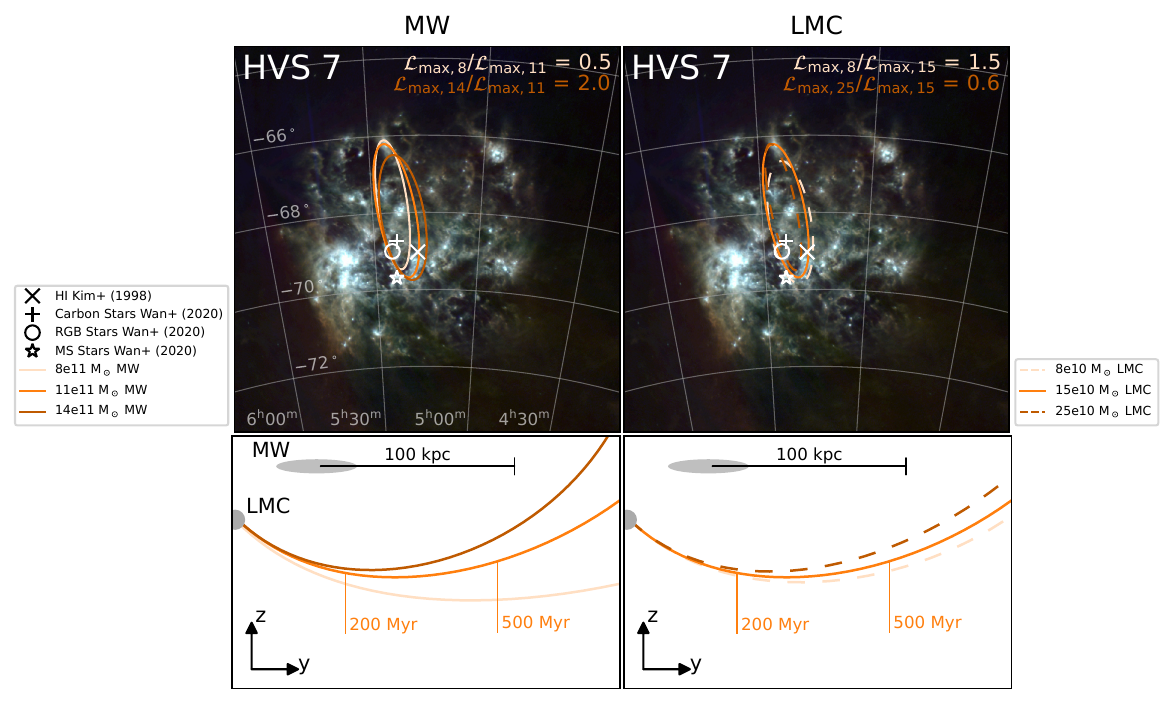}
    \caption{The changes in on-sky posterior and mean orbital trajectory for HVS 7 with different MW (left) and LMC (right) masses. The top panels show 1$\sigma$ ellipses for different masses, while the bottom panels show the mean of the orbits with posteriors greater than the 1$\sigma$ level. While the posterior distributions don't change significantly, there is strong variation in orbital path when changing the MW mass, and the LMC mass has a weaker effect. The bottom panels have the same extent as Figure~\ref{fig:orbits}.}
    \label{fig:masses}
\end{figure*}

In addition to our fiducial runs, we also ran smaller integrations with different MW and LMC masses. These integrations included 2,000 LMC orbits, each with 1,000 samples of HVS orbits. The MW masses used were $8\times10^{11}$~\msun\ and $1.4\times10^{12}$~\msun, the 1$\sigma$ values from \citet{bland-hawthorn16}. For the LMC, we used values of $8\times10^{10}$ and $2.5\times10^{11}$~\msun, consistent with previous works \citep{garavito-camargo19,Patel2020}. The top panels in Figure~\ref{fig:masses} show the 1$\sigma$ contours for HVS~7 after varying the galaxy masses. While there are minor changes, the contours are all consistent with each other, and the peak likelihood values are within 50\%.

However, these changes can have a significant impact on the morphology of the LMC's orbit around the MW. For more massive MWs, the LMC is more likely to have a shorter orbital period. This is shown in the bottom panels of Figure~\ref{fig:masses}. This has been studied in detail in previous works \citep[e.g.][]{besla07,kallivayalil13}, and we find similar results. Thus, the total masses of the galaxies remain an obstacle for a full reconstruction of the orbital history of the LMC and SMC.

\section{Choice of Prior} \label{app:prior}

Our fiducial analysis included Gaussian priors on the present-day position of the LMC's dynamical center (see Section~\ref{sec:likelihood}). However, we also adopted a tophat prior across the face of the LMC to show that our results are independent of our choice of prior. The tophat prior is defined to be equal to 1 for all pixels within a radius of $\sim3.1^\circ=2.7$~kpc (centered on (RA, Dec) $=(79.2,-68.7)$ degrees) and zero outside. Additionally, we apply a tophat in distance, with values of 1 for distances between 45 and 55~kpc.

\begin{figure*}
    \centering
    \includegraphics[width=0.367\textwidth]{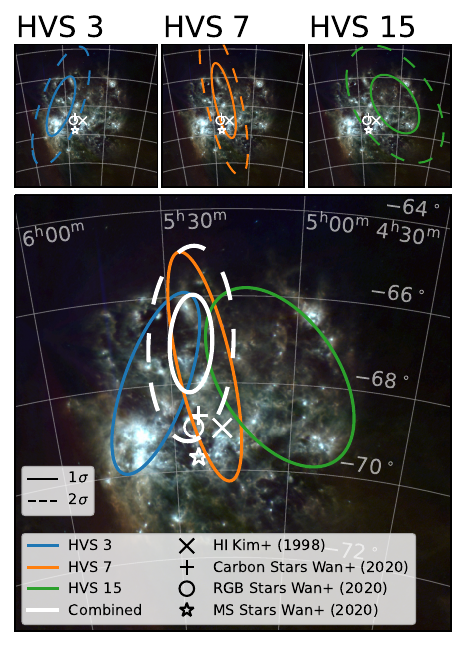}
    \includegraphics[width=0.55\textwidth]{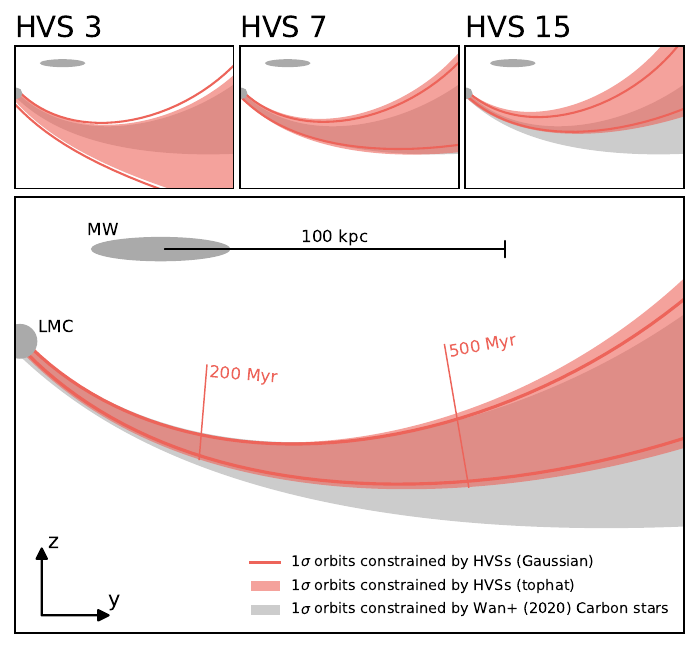}
    \caption{As Figures~\ref{fig:orbits} and \ref{fig:face-on} except with a tophat prior extending across the face of the LMC. On the right, the trajectories of the 1$\sigma$ orbits calculated with the tophat prior are shown as the red shaded regions, while the limits of the orbital space calculated with the Gaussian prior (see Figure~\ref{fig:orbits}) are shown as the red lines.}
    \label{fig:tophat}
\end{figure*}

As shown in the right panels of Figure~\ref{fig:tophat}, we see that the individual posterior distributions for each HVS are slightly larger, while the combined posterior is very similar to Figure~\ref{fig:face-on}. However the combined distribution is shifted slightly further away from the other dynamical center measurements. The combined Gaussian is located at (RA, Dec) $=(81.19,-67.08)$ degrees, with $\sigma_x=1.25^\circ$, $\sigma_y=1.12^\circ$, and position angle $\theta=2.2^\circ$.

This also results in a minor change to the orbital trajectories shown in the right panels of Figure~\ref{fig:tophat}.
The orbital space of the combined distribution is 65\% larger for the tophat prior than the Gaussian prior.
This shift then marginally includes the orbit from \citet{garavito-camargo21}.

\end{document}